\documentclass[12pt]{article}             
\usepackage{feynmf}

\newcommand{\seclabel}[1]{\label{#1}}
\newcommand{\figlabel}[1]{\label{#1}}
\newcommand{\bib}[1]{\bibitem{#1}}

\newcommand{\ia}{{\"{\i}}}
\newcommand{\absatz}{\vspace{2ex}\noindent}

\newcommand{\journal}[4]{{#1} {\bf{#2}}, #3 (#4)}
\newcommand{\JMathP}{\emph{J.\ Math.\ Phys.\ }}
\newcommand{\NPB}{\emph{Nucl.\ Phys.\ }{B}}
\newcommand{\PLB}{\emph{Phys.\  Lett.\ }{B}}
\newcommand{\PRD}{\emph{Phys.\ Rev.\ }{D}}
\newcommand{\book}[3]{\emph{#1}; #2 (#3)}
\newcommand{\preprint}[4][]{\emph{#2}; #3, #4 (to be published#1)}

\newcommand{\dis}{\displaystyle}
\newcommand{\non}{\nonumber}

\newcommand{\half}{\frac{1}{2}}
\newcommand{\e}{\mathrm{e}}
\newcommand{\ii}{\mathrm{i}}
\newcommand{\dd}{\mathrm{d}}

\newcommand{\xs}{\vec{X}_{\mathrm{s}}}
\newcommand{\ts}{T_{\mathrm{s}}}

\newcommand{\xu}{\vec{X}_{\mathrm{u}}}
\newcommand{\tu}{T_{\mathrm{u}}}

\newcommand{\Phis}{\Phi_{\mathrm{s}}}
\newcommand{\Phip}{\Phi_{\mathrm{p}}}
\newcommand{\phis}{\phi_{\mathrm{s}}}
\newcommand{\phip}{\phi_{\mathrm{p}}}
\newcommand{\As}{A_{\mathrm{s}}}
\newcommand{\Ap}{A_{\mathrm{p}}}
\newcommand{\Au}{A_{\mathrm{u}}}
\newcommand{\calAs}{\calA_{\mathrm{s}}}
\newcommand{\calAp}{\calA_{\mathrm{p}}}
\newcommand{\calAu}{\calA_{\mathrm{u}}}

\newcommand{\Xvs}{\vec{X}_{\mathrm{s}}}

\newcommand{\Ts}{T_{\mathrm{s}}}
\newcommand{\Tu}{T_{\mathrm{u}}}
\newcommand{\dedreiXs}{\dd^{3}\:\!\! X_{\mathrm{s}}\;}
\newcommand{\dedreiXu}{\dd^{3}\:\!\! X_{\mathrm{u}}\;}

\newcommand{\deTs}{\dd T_{\mathrm{s}}\;}
\newcommand{\deTu}{\dd T_{\mathrm{u}}\;}
\newcommand{\dek}{\dd k\;}
\newcommand{\dedk}{\frac{\dd^{d}\! k}{(2\pi)^d}\;}
\newcommand{\deint}[2]{\dd^{#1}\! #2\;}
\newcommand{\deintdim}[2]{\frac{\dd^{#1}\! #2}{(2\pi)^{#1}}\;}

\newcommand{\kv}{\vec{k}}
\newcommand{\pv}{\vec{\,\!p}\!\:{}}

\newcommand{\xv}{\vec{x}}

\newcommand{\de}{\partial}
\newcommand{\dev}{\vec{\de}}

\newcommand{\calA}{\mathcal{A}}\newcommand{\calL}{\mathcal{L}}
\newcommand{\calO}{\mathcal{O}}

\begin{document}

\begin{fmffile}{bsfeyn}
  \fmfset{curly_len}{2mm} \fmfset{wiggly_len}{3mm}
  \newcommand{\feynbox}[2]{\mbox{\parbox{#1}{#2}}}
\newcommand{\fs}{\scriptstyle} 
\newcommand{\hq}{\hspace{0.5em}}
  
%

\begin{titlepage}
\begin{flushright}
  hep-ph/9712467\\ NT@UW-98-3  \\ 18th December 1997 \\
\end{flushright}
\vspace*{1.5cm}
\begin{center}
  
  \LARGE{\textbf{Threshold Expansion and\\Dimensionally Regularised NRQCD}}

\end{center}
\vspace*{1.0cm}
\begin{center}
  \textbf{Harald W.\ Grie\3hammer\footnote{Email: hgrie@phys.washington.edu}}
  
  \vspace*{0.2cm}

  \emph{Nuclear Theory Group, Department of Physics, University of Washington,
    \\
    Box 351 560, Seattle, WA 98195-1560, USA} \vspace*{0.2cm}

\end{center}

\vspace*{2.0cm}


\begin{abstract}
  A Lagrangean and a set of Feynman rules are presented for non-relativistic
  QFT's with manifest power counting in the heavy particle velocity $v$. A
  r\'egime is identified in which energies and momenta are of order $Mv$. It is
  neither identical to the ultrasoft r\'egime corresponding to radiative
  processes with energies and momenta of order $Mv^2$, nor to the potential
  r\'egime with on shell heavy particles and Coulomb binding. In this soft
  r\'egime, massless particles are on shell, and heavy particle propagators
  become static. Examples show that it contributes to one- and two-loop
  corrections of scattering and production amplitudes near threshold. Hence,
  NRQFT agrees with the results of threshold expansion. A simple example also
  demonstrates the power of dimensional regularisation in NRQFT.
\end{abstract}

\vskip 1.0cm
\noindent
Suggested PACS numbers: 12.38.Bx, 12.39.Hg, 12.39.Jh.\\[1ex]
Suggested Keywords: non-relativistic QCD, effective field theory, threshold
expansion, \\
\phantom{Suggested Keywords:} dimensional regularisation.
\end{titlepage}

\setcounter{page}{2} \setcounter{footnote}{0} \newpage

%

\section{Introduction}
\seclabel{intro} \setcounter{equation}{0}

Velocity power counting in Non-Relativistic Quantum Field Theories (NRQFT)
\cite{CaswellLepage, BBL}, especially in NRQCD and NRQED, and identification of
the relevant energy and momentum r\'egimes has proven more difficult than
previously believed. In a recent article, Beneke and Smirnov
\cite{BenekeSmirnov} pointed out that the velocity rescaling rules proposed by
Luke and Manohar for Coulomb interactions~\cite{LukeManohar}, and by Grinstein
and Rothstein for bremsstrahlung processes~\cite{GrinsteinRothstein}, as united
by Luke and Savage~\cite{LukeSavage}, and by Labelle's power counting scheme in
time ordered perturbation theory~\cite{Labelle}, do not reproduce the correct
behaviour of the two gluon exchange contribution to Coulomb scattering between
non-relativistic particles near threshold. This has cast some doubt whether
NRQCD, especially in its dimensionally regularised version \cite{LukeSavage},
can be formulated using a self-consistent low energy Lagrangean. The aim of
this article is to demonstrate that a Lagrangean establishing explicit velocity
power counting exists, and to show that this Lagrangean reproduces the results
in Ref.\ \cite{BenekeSmirnov}.

This article is confined to outlining the ideas to resolve the puzzle,
postponing more formal arguments, calculations and derivations to a future,
longer publication \cite{hgpub4} which will also deal with gauge theories and
exemplary calculations.  It is organised as follows: In Sect.\ 
\ref{philosophy}, the relevant r\'egimes of NRQFT are identified. A simple
example demonstrates the usefulness of dimensional regularisation in enabling
explicit velocity power counting. Sect.\ \ref{rescaling} proposes the rescaling
rules necessary for a Lagrangean with manifest velocity power counting. The
Feynman rules are given. Simple examples in Sect.\ \ref{bsexamples} establish
further the necessity of the new, soft r\'egime introduced in Sect.\ 
\ref{philosophy}. Summary and outlook conclude the article, together with an
appendix on split dimensional regularisation \cite{LeibbrandtWilliams}.

\section{Idea of Dimensionally Regularised NRQFT}
\seclabel{philosophy} \setcounter{equation}{0}

For the sake of simplicity, let us -- following \cite{BenekeSmirnov} -- deal
with the Lagrangean
\begin{equation}
  \label{rellagr}
  \calL=(\de_\mu\Phi_\mathrm{R})^\dagger(\de^\mu\Phi_\mathrm{R}) - M^2
  \Phi_\mathrm{R}^\dagger\Phi_\mathrm{R}+ 
  \half(\de_\mu A)(\de^\mu A) - 2 M g\,\Phi_\mathrm{R}^\dagger\Phi_\mathrm{R} A
\end{equation}
of a heavy, complex scalar field $\Phi_\mathrm{R}$ with mass $M$ coupled to a
massless, real scalar $ A $. The coupling constant $g$ has been chosen
dimensionless. $\Phi_\mathrm{R}$ will be referred to as ``quark'' and $ A $ as
``gluon'' in a slight but clarifying abuse of language.  In NRQFT, excitations
with four-momenta bigger than $M$ are integrated out, giving rise to four-point
interactions between quarks. The first terms of the non-relativistically
reduced Lagrangean read
\begin{equation}
  \label{nrlagr}
  \calL_\mathrm{NRQFT}=\Phi^\dagger\Big(\ii\de_0+\frac{\dev^2}{2M} - g c_1\;A
  \Big)\Phi + \half(\de_\mu A)(\de^\mu A) + c_2\Big(\Phi^\dagger\Phi\Big)^2 +
  \dots\;\;,
\end{equation}
where the non-relativistic quark field is $\Phi=\sqrt{2M}\,\e^{\ii Mt}
\Phi_\mathrm{R}$ and the coefficients $c_i$ are to be determined by matching
relativistic and non-relativistic scattering amplitudes. To lowest order,
$c_1=1$ and $c_2=\frac{-g^4}{24\pi^2 M^2}$. The non-relativistic propagators
are
\begin{equation}
  \label{nonrelprop}
  \Phi\;:\;\frac{\ii}{T-\frac{\pv^2}{2M}+\ii\epsilon}\;\;,\;\;
   A\;:\;\frac{\ii}{k^2+\ii\epsilon}\;\;,
\end{equation}
where $T=p_0-M=\frac{\pv^2}{2M}+\dots$ is the kinetic energy of the quark.

When a Coulombic bound state of two quarks exists, the two typical energy and
momentum scales in the non-relativistic system are the bound state energy
$Mv^2$ and the relative momentum of the two quarks $Mv$ (i.e.\ the inverse size
of the bound state) \cite{BBL}. Here, $v=\beta\gamma\ll 1$ is the relativistic
generalisation of the relative particle velocity.  Cuts and poles in scattering
amplitudes close to threshold stem from bound states and on-shell propagation
of particles in intermediate states. They give rise to infrared divergences,
and in general dominate contributions to scattering amplitudes.  With the two
scales at hand, and energies and momenta being of either scale, three r\'egimes
are identified in which either $\Phi$ or $A$ in (\ref{nonrelprop}) is on shell:
\begin{eqnarray}
   \mbox{soft r\'egime: }&A_\mathrm{s}:&k_0\sim |\kv|\sim Mv\;\;,\non\\
  \label{regimes}
   \mbox{potential r\'egime: }&\Phi_\mathrm{p}:&T\sim Mv^2\;,\; |\pv|\sim
   Mv\;\;,\\
   \mbox{ultrasoft r\'egime: }&A_\mathrm{u}:&k_0\sim |\kv|\sim Mv^2\non
\end{eqnarray}
Ultrasoft gluons $\Au$ are emitted as bremsstrahlung or from excited states in
the bound system. Soft gluons $\As$ do not describe bremsstrahlung: Because in-
and outgoing quarks $\Phi_\mathrm{p}$ are close to their mass shell, they have
an energy of order $Mv^2$. Therefore, overall energy conservation forbids all
processes with outgoing soft gluons but without ingoing ones, and vice versa,
as their energy is of order $Mv$.

The list of particles is not yet complete: In a bound system, one needs gluons
which change the quark momenta but keep them close to their mass shell:
\begin{equation}
  \label{pgluon}
  A_\mathrm{p}\;\;:\;\;k_0\sim Mv^2\;,\;|\kv|\sim Mv
\end{equation}
So far, only potential gluons and quarks, and ultrasoft gluons had been
identified in the literature of power counting in NRQFT
\cite{LukeManohar,GrinsteinRothstein,Labelle}. That the soft r\'egime was
overlooked cast doubts on the completeness of NRQFT after Beneke and Smirnov
\cite{BenekeSmirnov} demonstrated its relevance near threshold in explicit one-
and two-loop calculations. In this article, the fields representing a
non-relativistic quark or gluon came naturally by identifying all possible
particle poles in the non-relativistic propagators, given the two scales at
hand.

When a soft gluon $A_\mathrm{s}$ couples to a potential quark $
\Phi_\mathrm{p}$, the outgoing quark is far off its mass shell and carries
energy and momentum of order $Mv$. Therefore, consistency requires the
existence of quarks in the soft r\'egime as well,
\begin{equation}
  \label{squark}
  \Phi_\mathrm{s}\;\;:\;\;T\sim |\pv|\sim Mv\;\;.
\end{equation}
As the potential quark has a much smaller energy than either of the soft
particles, it can -- by the uncertainty relation -- not resolve the precise
time at which the soft quark emits or absorbs the soft gluon. So, we expect a
``temporal'' multipole expansion to be associated with this vertex. In general,
the coupling between particles of different r\'egimes will not be point-like
but will contain multipole expansions for the particle belonging to the weaker
kinematic r\'egime. For the coupling of potential quarks to ultrasoft gluons,
this has been observed in Refs.\ \cite{GrinsteinRothstein,Labelle}.

Propagators will also be different from r\'egime to r\'egime: For soft quarks,
$\frac{\vec{p}^2}{2M}$ is negligible against the kinetic energy $T$, so that
the soft quark propagator may be expanded in powers of $\frac{\vec{p}^2}{2M}$,
and $\Phis$ is expected to become static to lowest order. As the energy of
potential gluons is much smaller than their momentum, the $\Ap$-propagator is
expected to become instantaneous.

With these five fields $\Phi_\mathrm{s},\;\Phi_\mathrm{p},\; A_\mathrm{s},
\;A_\mathrm{p},\;A_\mathrm{u}$ representing quarks and gluons in the three
different non-relativistic r\'egimes, soft, potential and ultrasoft, NRQFT
becomes self-consistent. The application of these ideas to NRQCD with the
inclusion of fermions and gauge particles is straightforward and will be
summarised in the next publication \cite{hgpub4}. An ultrasoft quark (which
would have a static propagator) is not relevant for this paper. It is hence not
considered, as is a fourth (``exceptional'') r\'egime in which momenta are of
the order $Mv^2$ and energies of the order $Mv$ or any r\'egime in which one of
the scales is set by $M$. They do not derive from poles in propagators, and
hence will be relevant only under ``exceptional'' circumstances. A future
publication \cite{hgpub4} has to prove that the particle content outlined is
not only consistent but complete.

It is worth noticing that the particles of the soft r\'egime can neither be
mimicked by potential gluon exchange, nor by contact terms generated by
integrating out the ultraviolet modes: Fields in the soft r\'egime have momenta
of the same order as the momenta of the potential r\'egime, but much higher
energies. Therefore, seen from the potential scale they describe instantaneous
but non-local interactions, as pointed out in \cite{BenekeSmirnov}. Integrating
out the scale $Mv$, one arrives at soft gluons and quarks as point-like
multi-quark interactions in the ultrasoft r\'egime. The physics of potential
quarks and gluons will still have to be described by spatially local, but
non-instantaneous interactions. The resulting theory -- baptised potential
NRQCD by Pineda and Soto \cite{PinedaSoto} -- can be derived from NRQCD as
presented here by integrating out the fields $\Phi_\mathrm{s},\;A_\mathrm{s}$
and $A_\mathrm{p}$.  There is no overlap between interactions and particles in
different r\'egimes.

\absatz In order to clarify this point, and before investigating the
interactions of the various r\'egimes further, the following example will
demonstrate the power of dimensional regularisation in NRQFT. It also
highlights some points which simplify the discussion of the following sections.
The integral corresponding to a one-dimensional loop
\begin{equation}
  \label{example1}
  I(a,b):=\int \dek \frac{1}{k^2-a^2+\ii\epsilon}\;\frac{1}{k^2-b^2+\ii
     \epsilon}=\frac{\ii\pi}{ab(a+b)}
\end{equation}
is easily calculated using contour integration. Assuming
$v^2:=\frac{a^2}{b^2}\ll 1$, the dominating contributions come from the regions
where $|k|$ is close to $a$ (``smaller r\'egime'') or $b$ (``larger
r\'egime''). Then, one can approximate the integral by
\begin{equation}
  \label{example2}
  I(a,b)\approx \bigg[\int\limits_{|k|\sim a} +\int\limits_{|k|\sim b}\bigg]
 \dek\frac{1}{k^2-a^2+\ii\epsilon}\; \frac{1}{k^2-b^2+\ii\epsilon} \;\;.
\end{equation}
In the first integral, $k$ is small against $b$, so that a Taylor expansion in
$\frac{k}{b}\sim v$ in that r\'egime is applicable and yields
\begin{equation}
  \label{example3}
  \frac{-1}{b^2}\sum\limits_{n=0}^\infty \int\limits_{|k|\sim
  a}\dek\frac{1}{k^2-a^2+\ii\epsilon}\; \frac{k^{2n}}{b^{2n}}\;\;.
\end{equation}
If $k^2$ becomes comparable to $b^2$, the expansion breaks down, so that the
approximated integral cannot be solved by contour integration. In general, the
(arbitrary) borders of the integration r\'egimes (the ``cutoffs'') will enter
in the result, and lead to divergences as they are taken to infinity because of
contributions from regions where $|k|\sim b \gg a$. A cutoff regularisation may
hence jeopardise power counting in $v$.

Dimensional regularisation overcomes this problem in a natural and elegant way:
If one treats (\ref{example3}) as a $d$-dimensional integral with $d\to1$ only
at the end of the calculation, the exact result will emerge as a power series
in $v=\frac{a}{b}$.  First, one extends the integration r\'egime from the
neighbourhood of $|a|$ to the whole $d$-dimensional space.  Then, one
calculates the integral order by order in the expansion, still treating
$\frac{k^2}{b^2}\sim v^2$ as formally small. Rewriting
\begin{equation}
  \label{example4}
  k^{2n}=\sum\limits_{m=0}^{n}{n \choose m} a^{2m} (k^2-a^2)^{n-m}\;\;,
\end{equation}
only the ($m=n$)-term contributes thanks to the fact that dimensionally
regularised integrals vanish when no intrinsic scale is present,
\begin{equation}
  \label{example5}
  \int\dedk k^\alpha=0\;\;.
\end{equation}
The result,
\begin{equation}
  \label{example6}
  \frac{\ii\pi}{a b^2}\sum\limits_{n=0}^{\infty}\frac{a^{2n}}{b^{2n}}
  \;\;\bigg(=\frac{\ii\pi}{a}\;\frac{1}{b^2-a^2}\bigg)
\end{equation}
is exactly the contribution one obtains in the contour integration from the
pole at $|k|=a$. Albeit the integral was expanded over the whole space,
dimensional regularisation missed the poles at $\pm b$ after expansion.

The integration about $|k|\sim b$ is treated likewise by expansion and
term-by-term dimensional regularisation. Adding this contribution,
\begin{equation}
  \label{example7}
  \frac{-\ii\pi}{b^3}\sum\limits_{n=0}^\infty \frac{a^{2n}}{b^{2n}} \;\;\bigg(=
  \frac{-\ii\pi}{b}\;\frac{1}{b^2-a^2}\bigg)\;\;,
\end{equation}
to (\ref{example6}), one obtains term by term the Taylor expansion of the exact
result (\ref{example1}) in the small parameter $v=\frac{a}{b}$.  Each of the
two regularised integrals sees only the pole in either of the r\'egimes
$|k|\sim a$ or $|k|\sim b$. Indeed, the overlap of the two r\'egimes is zero in
dimensional regularisation, even for arbitrary $v$. But then, the expansion in
the two different r\'egimes can be terminated only at the cost of low accuracy.

One could therefore have started with the definition of two different
integration variables, one formally living in the smaller r\'egime with
$|K_a|\sim a\sim vb$, the other formally living in the larger r\'egime,
$|K_b|\sim b$:
\begin{equation}
  \label{example8}
  \int\dek \to \int \deint{d}{K_a} + \int \deint{d}{K_b}
\end{equation}
The momentum $k$ is represented in each of the kinematic r\'egimes by either
$K_a$ or $K_b$.  The integrands \emph{must} then be expanded in a formal way
\emph{as if} $|K_a|\sim a\sim vb$ and $|K_b|\sim b$. Otherwise, the poles are
double counted.  If one wants to calculate to a certain order in $v$, the
expansion in the different variables $\frac{K_a}{b}$ and $\frac{a}{K_b}$ is
just terminated at the appropriate order. No double counting of the poles can
occur. Coming back to the three different r\'egimes of NRQFT (\ref{regimes}),
there will therefore be no double counting between any pair of domains.

Finally, note that the limit $a\to 0$ is not smooth: For $a=0$, dimensional
regularisation of (\ref{example3}) is zero because of the absence of a scale
(\ref{example5}). A pinch singularity encountered in contour integration at
$k=\pm \ii \epsilon$ behaves hence like a pole of second order in dimensional
regularisation and is discarded, see also App.\ \ref{app:splitdimreg}.

By induction, the arguments presented here can be extended to prove that for
any convergent one-dimensional integral containing several scales, Laurent
expansion about each saddle point and dimensional regularisation gives the same
result as contour integration. A formal proof of the validity of threshold
expansion does presently not exist for the case of multi-dimensional and
divergent integrals, but Beneke and Smirnov~\cite{BenekeSmirnov} could
reproduce the correct structures of known non-trivial two-loop integrals using
threshold expansion, which is highly suggestive that such a proof can be given.
This claim is supported by the observation that threshold expansion is very
similar to the asymptotic expansion of dimensionally regularised integrals in
the limit of loop momenta going to infinity, for which such a proof
exists~\cite{Smirnov}~\footnote{I am indebted to M.\ Beneke for conversation on
  this point}.

\section{Rescaling Rules, Lagrangean and Feynman Rules}
\seclabel{rescaling} \setcounter{equation}{0}

In order to establish explicit velocity power counting in the NRQFT Lagrangean,
one rescales the space-time coordinates such that typical momenta in either
r\'egime become dimensionless, as first proposed in \cite{LukeManohar} for the
potential r\'egime, and in \cite{GrinsteinRothstein} for the ultrasoft
r\'egime:
\begin{eqnarray}
  \mbox{soft: } && t=(Mv)^{-1} \;\ts\;\;,\;\; \xv=(Mv)^{-1}\;\xs\;\;,\non\\
  \label{xtscaling}
  \mbox{potential: }&& t=(Mv^2)^{-1}\; \tu\;\;,\;\; \xv=(Mv)^{-1}\;\xs\;\;,\\
\mbox{ultrasoft: }&& t=(Mv^2)^{-1}\; \tu\;\;,\;\;\xv=(Mv^2)^{-1}\;\xu\;\;.\non
\end{eqnarray}
In order for the propagator terms in the NRQFT Lagrangean to be properly
normalised, one has to set for the representatives of the gluons in the three
r\'egimes
\begin{eqnarray}
   \mbox{soft: } && \As(\xv,t) = (Mv)\; \calAs(\xs,\ts)\;\;,\non\\
   \label{gluonscaling}
   \mbox{potential: } && \Ap(\xv,t) =(Mv^{\frac{3}{2}})\;\calAp(\xs,\tu)\;\;,\\
   \mbox{ultrasoft: } && \Au(\xv,t) = (Mv^2)\; \calAu(\xu,\tu)\;\;,\non
\end{eqnarray}
and for the quark representatives
\begin{eqnarray}
  \label{quarkscaling}
  \mbox{soft: } && \Phi_\mathrm{s}(\xv,t) = (Mv)^{\frac{3}{2}}\;
  \phi_\mathrm{s} (\xs,\ts)\;\;,\\
  \mbox{potential: } &&\Phi_\mathrm{p}(\xv,t) = (Mv)^{\frac{3}{2}}\;
  \phi_\mathrm{p}(\xs,\tu)\;\;.\non
\end{eqnarray}
The rescaled free Lagrangean in the three regions reads then
\begin{eqnarray}
  \label{sfreelagr}
   \mbox{soft: } &&\dedreiXs\deTs\Big[\phi_\mathrm{s}^\dagger\Big(\ii\de_0+
   \frac{v}{2} \vec{\de}^2 \Big)\phi_\mathrm{s} + \half(\de_\mu
   \calA_\mathrm{s})(\de^\mu \calA_\mathrm{s})\Big]\;\;,\\
  \label{pfreelagr}
   \mbox{potential: } &&\dedreiXs\deTu\Big[
   \phi_\mathrm{p}^\dagger\Big(\ii\de_0+\frac{1}{2} \vec{\de}^2 \Big)
   \phi_\mathrm{p} + \half\Big(\calA_\mathrm{p} \dev^2 \calA_\mathrm{p} - v^2
   \calA_\mathrm{p} \de^2_0 \calA_\mathrm{p}\Big) \Big]\;\;,\\
  \label{ufreelagr}
   \mbox{ultrasoft: } && \dedreiXu\deTu \half(\de_\mu \calAu)(\de^\mu
\calAu)\;\;.
\end{eqnarray}
Here, as in the following, the positions of the fields have been left out
whenever they coincide with the variables of the volume element.  Derivatives
are to be taken with respect to the rescaled variables of the volume element.
The (un-rescaled) propagators are
\begin{eqnarray}
  \label{sprop}
       \mbox{soft: } && \Phis:\; \feynbox{60\unitlength}{
    \begin{fmfgraph*}(60,30)
      \fmfleft{i}\fmfright{o}
      \fmf{heavy,width=thin,label=$\fs(T,,\pv)$,label.side=left}{i,o}
    \end{fmfgraph*}}
  \;=\;\frac{\ii}{T+\ii\epsilon}\;\;,\\
  && \As:\;
    \feynbox{60\unitlength}{
      \begin{fmfgraph*}(60,30)
        \fmfleft{i}\fmfright{o} \fmf{zigzag,width=thin,label=$\fs
          k$,label.side=left}{i,o}
    \end{fmfgraph*}}
  \;=\;\frac{\ii}{k^2+\ii\epsilon}\;\;,\non\\
  \label{pprop}
    \mbox{potential: } && \Phip:\;
    \feynbox{60\unitlength}{
    \begin{fmfgraph*}(60,30)
      \fmfleft{i}\fmfright{o}
      \fmf{fermion,width=thick,label=$\fs(T,,\pv)$,label.side=left}{i,o}
    \end{fmfgraph*}}
  \;=\; \frac{\ii}{T-\frac{\pv^2}{2M}+\ii\epsilon}\;\;,\\
  &&
    \Ap:\; \feynbox{60\unitlength}{
      \begin{fmfgraph*}(60,30)
        \fmfleft{i}\fmfright{o} \fmf{dashes,width=thin,label=$\fs k$
          ,label.side=left}{i,o}
   \end{fmfgraph*}}
    \;=\;\frac{\ii}{-\kv^2+\ii\epsilon}\;\;,\non\\
    \label{uprop} \mbox{ultrasoft: }
    && \Au:\; \feynbox{60\unitlength}{
      \begin{fmfgraph*}(60,30)
        \fmfleft{i}\fmfright{o} \fmf{photon,width=thin,label=$\fs k$
          ,label.side=left}{i,o}
   \end{fmfgraph*}}
   \;=\;\frac{\ii}{k^2+\ii\epsilon}\;\;.
\end{eqnarray}
As expected, the soft quark becomes static, resembling the quark propagator in
heavy quark effective theory, and the potential gluon becomes instantaneous. In
order to maintain velocity power counting, corrections of order $v$ or higher
must be treated as insertions as in the example, (\ref{example1}). Insertions
are represented by the (un-rescaled) Feynman rules
\begin{equation}
  \label{insertions}
      \feynbox{60\unitlength}{
        \begin{fmfgraph*}(60,30)
          \fmfleft{i}\fmfright{o} \fmf{double,width=thin}{i,v,o}
          \fmfv{decor.shape=cross,label=$\fs(T,,\pv)$, label.angle=90}{v}
   \end{fmfgraph*}}
   \;=\;-\ii\;\frac{\pv^2}{2M}\;=\;\calO(v)
    \;\;,\;\;\feynbox{60\unitlength}{
      \begin{fmfgraph*}(60,30)
        \fmfleft{i}\fmfright{o} \fmf{dashes,width=thin}{i,v,o}
        \fmfv{decor.shape=cross,label=$\fs k$,label.angle=90}{v}
  \end{fmfgraph*}}
    \;=\;+ \ii k_0^2\;=\;\calO(v^2) \;\;.
\end{equation}
Except for the physical gluons $\As$ and $\Au$, there is no distinction between
Feynman and retarded propagators in NRQFT: Antiparticle propagation has been
eliminated by the field transformation from the relativistic to the
non-relativistic Lagrangean, and both propagators have maximal support for
on-shell particles, the Feynman propagator outside the light cone vanishing
like $\e^{-M}$. Feynman's perturbation theory becomes more convenient than the
time-ordered formalism, as less diagrams have to be calculated .

Finally, the interaction part of the Lagrangean reads (neglecting for the
moment the $\Phi^4$ vertex in (\ref{nrlagr}))
\begin{eqnarray}
  \label{sintlagr}
       \mbox{soft: } && \dedreiXs\deTs (-g)\Big[\Big(\calAs+
    \sqrt{v}\;\calAp(\Xvs,v\Ts)+v\;\calAu(v\Xvs,v\Ts)\Big) \phis^\dagger\phis +
    \\&&\phantom{\dedreiXs\deTs (-g)\Big[}
    +\Big(\calAs\phis^\dagger\phip(\Xvs,v\Ts) +\mathrm{ h.c. } \Big)\Big]
    \non\\
    \label{pintlagr} \mbox{potential: } && \dedreiXs\deTu
    (-g)\Big(\frac{1}{\sqrt{v}}\calAp + \calAu(v\Xvs,\Tu)\Big)
    \phip^\dagger\phip\;\;.
\end{eqnarray}
Note that the scaling r\'egime of the volume element is set by the particle
with the highest momentum and energy. Vertices like $\calAs\phip^\dagger\phip$
cannot occur as energy and momentum must be conserved within each r\'egime to
the order in $v$ one works.  Amongst the fields introduced, these are the only
interactions within and between different r\'egimes allowed.  One sees that
technically, the multipole expansion comes from the different scaling of $\xv$
and $t$ in the three r\'egimes. It is also interesting to note that there is no
choice but to assign one and the same coupling strength $g$ to each
interaction. Different couplings for one vertex in different r\'egimes are not
allowed. This is to be expected, as the example (\ref{example8}) demonstrated
that the fields in the various r\'egimes are representatives of one and the
same non-relativistic particle, whose interactions are fixed by the
non-relativistic Lagrangean (\ref{nrlagr}).
  
The interaction Feynman rules are \vspace*{2ex}
\begin{eqnarray}
      \label{pppvertex} \feynbox{55\unitlength}{
        \begin{fmfgraph*}(55,30)
          \fmfstraight \fmftop{i,o} \fmfbottom{u}
          \fmf{fermion,width=thick,label=$\fs(T,,\pv)$,label.side=left}{i,v}
          \fmf{fermion,width=thick,label=$\fs(T^\prime,,\pv^\prime)$,
            label.side=right}{o,v} \fmffreeze
          \fmf{dashes,width=thin,label=$\uparrow \fs k$}{u,v} \end{fmfgraph*}}
      &\!\!\!\!=&\!\! -\ii g (2\pi)^4 \;\delta(T+T^\prime+k_0)
      \;\delta^{(3)}(\pv+\pv^\prime+\kv)\;=\;\calO(\frac{1}{\sqrt{v}}) \;\;,
      \\[3ex]
    \label{pupvertex} \feynbox{55\unitlength}{
      \begin{fmfgraph*}(55,30)
        \fmfstraight \fmftop{i,o} \fmfbottom{u}
    \fmf{fermion,width=thick
      }{i,v}
    \fmf{fermion,width=thick
    }{o,v} \fmffreeze \fmf{photon,width=thin
    }{u,v}
    \end{fmfgraph*}} &\!\!\!\!=&\!\! -\ii g (2\pi)^4 \;\delta(T+T^\prime+k_0)
    \Big[\exp\Big({\kv\cdot\frac{\partial}{\partial(\pv+\pv^\prime)}}\Big)
    \;\delta^{(3)}(\pv+\pv^\prime)\Big]\;=\;\calO(\e^v)\;\;, \\[3ex]
    \label{sssvertex}
    \feynbox{55\unitlength}{
      \begin{fmfgraph*}(55,30) \fmfstraight \fmftop{i,o}
    \fmfbottom{u} \fmf{heavy,width=thin
      }{i,v}
    \fmf{heavy,width=thin
    }{o,v} \fmffreeze \fmf{zigzag,width=thin
    }{u,v}
    \end{fmfgraph*}} &\!\!\!\!=&\!\! -\ii g (2\pi)^4 \;\delta(T+T^\prime+k_0)
    \;\delta^{(3)}(\pv+\pv^\prime+\kv)\;=\;\calO(v^0) \;\;, \\[3ex]
    \label{sspvertex}
    \feynbox{55\unitlength}{ \begin{fmfgraph*}(55,30) \fmfstraight \fmftop{i,o}
    \fmfbottom{u} \fmf{heavy,width=thin
      }{i,v}
    \fmf{fermion,width=thick
    }{o,v} \fmffreeze \fmf{zigzag,width=thin
    }{u,v}
    \end{fmfgraph*}} &\!\!\!\!=&\!\! -\ii g (2\pi)^4
    \Big[\exp\Big({T^\prime\;\frac{\partial}{\partial (T+k_0)}}\Big)
    \;\delta(T+k_0)\Big] \;\delta^{(3)}(\pv+\pv^\prime+\kv)\;=\;\calO(\e^v)
    \;\;, \\[3ex]
    \label{spsvertex} \feynbox{55\unitlength}{ \begin{fmfgraph*}(55,30)
        \fmfstraight \fmftop{i,o} \fmfbottom{u}
    \fmf{heavy,width=thin
      }{i,v}
    \fmf{heavy,width=thin
    }{o,v} \fmffreeze \fmf{dashes,width=thin
    }{u,v}
    \end{fmfgraph*}} &\!\!\!\!=&\!\! -\ii g (2\pi)^4
    \Big[\exp\Big({k_0\;\frac{\partial}{\partial (T+T^\prime)}}\Big)
    \;\delta(T+T^\prime)\Big]
    \;\delta^{(3)}(\pv+\pv^\prime+\kv)\,=\,\calO(\sqrt{v}\;\e^v), \\[3ex]
    \label{susvertex} \feynbox{55\unitlength}{ \begin{fmfgraph*}(55,30)
    \fmfstraight \fmftop{i,o} \fmfbottom{u}
    \fmf{heavy,width=thin
      }{i,v}
    \fmf{heavy,width=thin
    }{o,v} \fmffreeze \fmf{photon,width=thin
    }{u,v}
    \end{fmfgraph*}}
     &\!\!\!\!=&\!\! -\ii g (2\pi)^4
    \Big[\exp\Big({k_0\;\frac{\partial}{\partial (T+T^\prime)}}\Big)
    \;\delta(T+T^\prime)\Big]\times\\ \non &&\phantom{-\ii g (2\pi)^4}\times
    \Big[\exp\Big({\kv\cdot\frac{\partial}{\partial(\pv+\pv^\prime)}}\Big)
    \;\delta^{(3)}(\pv+\pv^\prime)\Big]\;=\;\calO(v\;\e^v)\;\;.
\end{eqnarray}
The exponents representing the multipole expansion have to be expanded to the
desired order in $v$. Double counting is prevented by the fact that in addition
to most of the propagators, all vertices are distinct because of different
multipole expansions.

Using the equations of motion, the temporal multipole expansion may be
re-written such that energy becomes conserved at the vertex. Now, both soft and
potential or ultrasoft energies are present in the propagators, making it
necessary to expand it in ultrasoft and potential energies. An example would be
to restate the vertex (\ref{spsvertex}) as
\begin{eqnarray}
      \feynbox{55\unitlength}{
  \begin{fmfgraph*}(55,20)
    \fmfstraight \fmftop{i,o} \fmfbottom{u}
     \fmf{heavy,width=thin
       }{i,v}
     \fmf{heavy,width=thin
       }{o,v} \fmffreeze
     \fmf{dashes,width=thin
       }{u,v}
   \end{fmfgraph*}}
   &=& -\ii g (2\pi)^4 \;\delta(T+T^\prime+k_{\mathrm{p},0})
   \;\delta^{(3)}(\pv+\pv^\prime+\kv)\;=\;\calO(\sqrt{v}) \;\;,
\end{eqnarray}
and the soft propagator as containing insertions $\calO(v)$ for potential
energies $k_{\mathrm{p}}$
\begin{equation}
  \feynbox{55\unitlength}{
    \begin{fmfgraph*}(55,30)
      \fmfleft{i}\fmfright{o} \fmf{heavy,width=thin,label=
           $\fs(T+k_{\mathrm{p},,0},,\pv)$,label.side=left}{i,o}
    \end{fmfgraph*}}
        \;=\;\frac{\ii}{T+\ii\epsilon}
    \sum\limits_{n=0}^\infty\left(\frac{-k_{\mathrm{p},0} }{T}\right)^n \;\;.
\end{equation}  
The same holds of course for the momentum-non-conserving vertices.

In the renormalisation group approach, there is therefore only one relevant
coupling (i.e.\ only one which dominates at zero velocity): As expected, it is
the $\Phip\Phip\Ap$ coupling providing the binding. The $\Phis\Phis\Ap$-,
$\Phis\Phis\Au$-couplings and both insertions (\ref{insertions}) are
irrelevant. The marginal couplings $\Phip\Phip\Au$, $\Phis\Phis\As$ and
$\Phis\Phip\As$ are irrelevant in gauge theories in carefully chosen gauges
like the Coulomb gauge. This point will be elaborated upon in the future
\cite{hgpub4}.

The velocity power counting is not yet complete. As one sees from the volume
element used in (\ref{sintlagr}), the vertex rules for the soft r\'egime count
powers of $v$ with respect to the soft r\'egime. One hence retrieves the
velocity power counting of Heavy Quark Effective
Theory~\cite{IsgurWise1,IsgurWise2} (HQET), in which the interactions between
one heavy (and hence static) and one or several light quarks are described.
Usually, HQET counts inverse powers of mass in the Lagrangean, but because in
the soft r\'egime $Mv\sim \mbox{const.}$, the two approaches are actually
equivalent.  HQET becomes a sub-set of NRQCD, complemented by interactions
between soft (HQET) and potential or ultrasoft particles.

In NRQCD with two potential quarks as initial and final states, the soft
r\'egime can occur only inside loops, as noted above. Therefore, the power
counting in the soft sub-graph has to be transfered to the potential r\'egime.
Because soft loop momenta scale like $[\dd^{4}\! k_\mathrm{s}]\sim v^4$, while
potential ones like $[\dd^{4}\! k_\mathrm{p}]\sim v^5$, each largest sub-graph
which contains only soft quarks and no potential ones (a ``soft blob'') is
enhanced by an additional factor $\frac{1}{v}$. Couplings between soft quarks
and any gluons inside a blob take place in the soft r\'egime and hence are
counted according to the rules of that r\'egime. Each soft blob contributes at
least four orders of $g$, but only one inverse power of $v\sim g^2$. Power
counting is preserved.  These velocity power counting rules in loops are
verified in explicit calculations of the exemplary graphs (see also below), but
a rigorous derivation is left for a future publication~\cite{hgpub4}.

With rescaling, multipole expansion and loop counting, the velocity power
counting rules are established, and one can now proceed to check the validity
of the proposed Lagrangean, matching NRQFT to the relativistic theory in the
examples given by Beneke and Smirnov \cite{BenekeSmirnov}.

\section{Model Calculations}
\seclabel{bsexamples} \setcounter{equation}{0}

The first example is the lowest order correction to the two quark production
graph. Without proof, it will be used that in dimensional regularisation, one
can match NRQFT and the relativistic theory graph by graph, so that not the
whole scattering amplitude has to be considered \cite{BenekeSmirnov}. The
collection of graphs to be matched to the relativistic diagram is depicted in
Fig.\ \ref{graph1}.
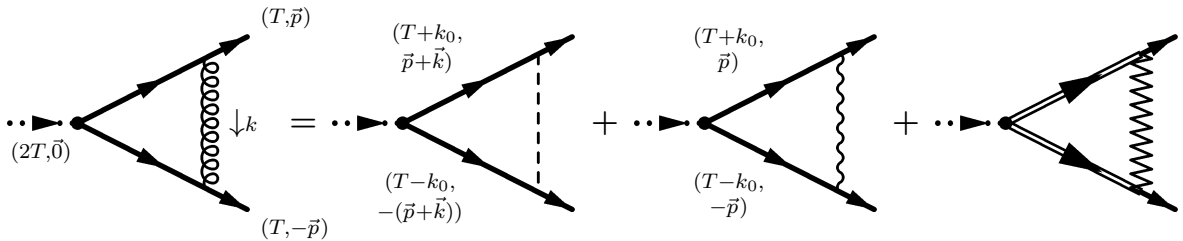
\begin{figure}[!htb]           
  \begin{center}
    \vspace*{2ex} \feynbox{100\unitlength}{ \begin{fmfgraph*}(90,65)
        \fmfstraight \fmfleft{i1} \fmfright{o2,o1}
        \fmf{ghost,width=thick,tension=2,label={$\fs(2T,,\vec{0})$}}{i1,v1}
        \fmfdot{v1} \fmf{fermion,width=thick,tension=0.5}{v1,v2}
        \fmf{fermion,width=thick,tension=0.5}{v1,v3}
        \fmf{fermion,width=thick,tension=2}{v3,o2}
        \fmf{fermion,width=thick,tension=2}{v2,o1} \fmflabel{$\fs(T,\pv)$}{o1}
        \fmflabel{$\fs(T,-\pv)$}{o2} \fmffreeze
        \fmf{gluon,width=thin,label=$\fs{\dis\downarrow}
          k$,label.side=left}{v2,v3}
    \end{fmfgraph*} }\hq\bf{=}\hq\feynbox{90\unitlength}{
    \begin{fmfgraph*}(90,65) \fmfstraight \fmfleft{i1} \fmfright{o2,o1}
      \fmf{ghost,width=thick,tension=2}{i1,v1} \fmfdot{v1}
      \fmf{fermion,width=thick,tension=0.5,label=${(T+k_0 ,, \atop
          \pv+\kv)}$,label.side=left}{v1,v2}
      \fmf{fermion,width=thick,tension=2}{v2,o1}
      \fmf{fermion,width=thick,tension=0.5,label=${(T-k_0,,\atop-(\pv+\kv))}$,
        label.side=right}{v1,v3} \fmf{fermion,width=thick,tension=2}{v3,o2}
      \fmffreeze \fmf{dashes,width=thin}{v2,v3} \end{fmfgraph*}
    }\hq\bf{+}\hq\feynbox{90\unitlength}{ \begin{fmfgraph*}(90,65) \fmfstraight
      \fmfleft{i1} \fmfright{o2,o1} \fmf{ghost,width=thick,tension=2}{i1,v1}
      \fmfdot{v1}
      \fmf{fermion,width=thick,tension=0.5,label=${(T+k_0,,\atop\pv)}$,
        label.side=left}{v1,v2} \fmf{fermion,width=thick,tension=2}{v2,o1}
      \fmf{fermion,width=thick,tension=0.5,label=${(T-k_0,,\atop-\pv)}$,
        label.side=right}{v1,v3} \fmf{fermion,width=thick,tension=2}{v3,o2}
      \fmffreeze \fmf{photon,width=thin}{v2,v3} \end{fmfgraph*}
    }\hq\bf{+}\hq\feynbox{90\unitlength}{ \begin{fmfgraph*}(90,65) \fmfstraight
      \fmfleft{i1} \fmfright{o2,o1} \fmf{ghost,width=thick,tension=2}{i1,v1}
      \fmfdot{v1} \fmf{heavy,width=thin,tension=0.5}{v1,v2}
      \fmf{fermion,width=thick,tension=2}{v2,o1}
      \fmf{heavy,width=thin,tension=0.5}{v1,v3}
      \fmf{fermion,width=thick,tension=2}{v3,o2} \fmffreeze
      \fmf{zigzag,width=thin}{v2,v3} \end{fmfgraph*} }
\end{center}
  \caption{\figlabel{graph1}\sl Matching of the $\calO(g^2)$ correction
    to two quark production off an external current to lowest order in $v$ in
    each of the three r\'egimes.}
\end{figure}         
Here and in the following, hard (ultraviolet) contributions will not be shown
explicitly. They are taken care of by the four-quark interaction of the
non-relativistic Lagrangean (\ref{nrlagr}) and renormalisation of the external
currents \cite{LukeSavage}.

The energy and momentum routing has been chosen to be the one of the
non-relativistic center of mass system, with $2T$ the total kinetic energy, and
$y=-(\pv)^2\propto - v^2$ the relative four-momentum squared of the outgoing
quarks as indicator for the thresholdness of the process considered. Thanks
again to dimensional regularisation, any other assignment can be chosen and
reproduces the result.

The vanishing of the ultrasoft gluon exchange diagram and the value of the
potential gluon exchange diagram have already been calculated in
\cite{LukeSavage}. The soft exchange diagram vanishes, so that no new
contribution is obtained. It is not even necessary to specify how soft quarks
couple to external sources: If energy is conserved at the production vertex,
the integral to be calculated is
\begin{equation}
  \label{vertex1}
  \int \dedk\frac{1}{T+k_0+\ii\epsilon}\;\frac{1}{T-k_0+\ii\epsilon}\;
  \frac{1}{k_0^2-\kv^2}\;\;.
\end{equation}
As the gluon is soft, $T\ll k_0$ and the quark propagators must be expanded in
$T/k_0\sim v$, giving zero to any order as no scale is present in the
dimensionally regularised integral. If energy is not conserved at the
production vertex, the soft quark propagator is $\frac{1}{\pm k_0}$, and the
contribution vanishes again. Therefore, there is no coupling of soft subgraphs
to external sources to any order in $v$. Soft quarks in external lines are far
off their mass shell and hence violate the assumptions underlying threshold
expansion and NRQFT. In general, we conclude that soft quarks are present only
in internal lines, and that the first non-vanishing contribution from the soft
r\'egime for the production vertex occurs not earlier than at $\calO (g^4)$.

The first soft non-zero contribution comes actually from the two gluon direct
exchange diagram of Fig.\ \ref{graph2} calculated by Beneke and Smirnov
\cite{BenekeSmirnov} using threshold expansion. The Mandelstam variable
$t=-(\pv-\pv^\prime)^2$ describes the momentum transfer in the center of mass
system.
\begin{figure}[!htb]
  \vspace*{2ex}
      \begin{center} \feynbox{120\unitlength}{
    \begin{fmfgraph*}(100,60) \fmfstraight \fmftop{i1,o1} \fmfbottom{i2,o2}
      \fmf{fermion,width=thick,tension=1,label=$\fs(T,,\pv)$,
        label.side=left}{i1,v1} \fmf{fermion,width=thick,tension=0.5}{v1,v2}
      \fmf{fermion,width=thick,tension=1,label=$\fs(T,,\pv^\prime)$,
        label.side=left}{v2,o1}
      \fmf{fermion,width=thick,tension=1,label=$\fs(T,,-\pv)$,
        label.side=right}{i2,v3} \fmf{fermion,width=thick,tension=0.5}{v3,v4}
      \fmf{fermion,width=thick,tension=1,label=$\fs(T,,-\pv^\prime)$,
        label.side=right}{v4,o2} \fmffreeze \fmf{gluon,width=thin,label=$\fs
        k{\dis\uparrow}$,label.side=left,tension=0.5}{v3,v1}
      \fmf{gluon,width=thin,label=${\dis\downarrow}{
          (k_0,,\atop\pv-\pv^\prime+\kv)}$,label.side=right,tension=0.5}{v4,v2}
    \end{fmfgraph*} }
  \hq\bf{=}\hq\feynbox{110\unitlength}{
    \begin{fmfgraph*}(100,60) \fmfstraight \fmftop{i1,o1} \fmfbottom{i2,o2}
      \fmf{fermion,width=thick,tension=1.5}{i1,v1}
      \fmf{fermion,width=thick,tension=0.5}{v1,v2}
      \fmf{fermion,width=thick,tension=1.5}{v2,o1}
      \fmf{fermion,width=thick,tension=1.5}{i2,v3}
      \fmf{fermion,width=thick,tension=0.5}{v3,v4}
      \fmf{fermion,width=thick,tension=1.5}{v4,o2} \fmffreeze
      \fmf{photon,width=thin,label=$\fs
        k{\dis\uparrow}$,label.side=left,tension=0.5}{v3,v1}
      \fmf{photon,width=thin,label=$\fs{\dis\downarrow}
        k$,label.side=right,tension=0.5}{v4,v2}
  \end{fmfgraph*}
  }\hq\bf{+}\hq\feynbox{115\unitlength}{
    \begin{fmfgraph*}(100,60)
      \fmfstraight \fmftop{i1,o1} \fmfbottom{i2,o2}
      \fmf{fermion,width=thick,tension=1.5}{i1,v1}
      \fmf{fermion,width=thick,tension=0.5}{v1,v2}
      \fmf{fermion,width=thick,tension=1.5}{v2,o1}
      \fmf{fermion,width=thick,tension=1.5}{i2,v3}
      \fmf{fermion,width=thick,tension=0.5}{v3,v4}
      \fmf{fermion,width=thick,tension=1.5}{v4,o2} \fmffreeze
      \fmf{photon,width=thin,label=$\fs
        k{\dis\uparrow}$,label.side=left,tension=0.5}{v3,v1}
      \fmf{dashes,width=thin,label=${\dis\downarrow}
        {(k_0,,\atop\pv^\prime-\pv)} $,label.side=right,tension=0.5}{v4,v2}
  \end{fmfgraph*} }\hq\bf{+}\hq

\vspace*{30\unitlength}

\bf{+}\hq\feynbox{100\unitlength}{
      \begin{fmfgraph*}(100,60) \fmfstraight
        \fmftop{i1,o1} \fmfbottom{i2,o2}
        \fmf{fermion,width=thick,tension=1.5}{i1,v1}
        \fmf{fermion,width=thick,tension=0.5}{v1,v2}
        \fmf{fermion,width=thick,tension=1.5}{v2,o1}
        \fmf{fermion,width=thick,tension=1.5}{i2,v3}
        \fmf{fermion,width=thick,tension=0.5}{v3,v4}
        \fmf{fermion,width=thick,tension=1.5}{v4,o2} \fmffreeze
        \fmf{dashes,width=thin,tension=0.5}{v3,v1}
        \fmf{photon,width=thin,tension=0.5}{v4,v2}
  \end{fmfgraph*}
  }\hq\bf{+}\hq\feynbox{100\unitlength}{
    \begin{fmfgraph*}(100,60)
      \fmfstraight \fmftop{i1,o1} \fmfbottom{i2,o2}
      \fmf{fermion,width=thick,tension=1.5}{i1,v1}
      \fmf{fermion,width=thick,tension=0.5}{v1,v2}
      \fmf{fermion,width=thick,tension=1.5}{v2,o1}
      \fmf{fermion,width=thick,tension=1.5}{i2,v3}
      \fmf{fermion,width=thick,tension=0.5}{v3,v4}
      \fmf{fermion,width=thick,tension=1.5}{v4,o2} \fmffreeze
      \fmf{dashes,width=thin,tension=0.5}{v3,v1}
      \fmf{dashes,width=thin,tension=0.5}{v4,v2} \end{fmfgraph*}
    }\hq\bf{+}\hq\feynbox{100\unitlength}{
    \begin{fmfgraph*}(100,60)
      \fmfstraight \fmftop{i1,o1} \fmfbottom{i2,o2}
      \fmf{fermion,width=thick,tension=1.5}{i1,v1}
      \fmf{heavy,width=thin,tension=0.5,label=$\fs(k_0,,\pv+\kv)$,
        label.side=left}{v1,v2} \fmf{fermion,width=thick,tension=1.5}{v2,o1}
      \fmf{fermion,width=thick,tension=1.5}{i2,v3}
      \fmf{heavy,width=thin,tension=0.5,label=$\fs (-k_0,,-\pv-\kv)$,
        label.side=right}{v3,v4} \fmf{fermion,width=thick,tension=1.5}{v4,o2}
      \fmffreeze \fmf{zigzag,width=thin,tension=0.5}{v3,v1}
      \fmf{zigzag,width=thin,tension=0.5}{v4,v2}
  \end{fmfgraph*} }
\end{center}
\caption{\figlabel{graph2}\sl Planar \protect$\calO(g^4)$ contributions to
  Coulomb scattering. The four-point interaction and insertion diagrams are not
  displayed.}
\end{figure}
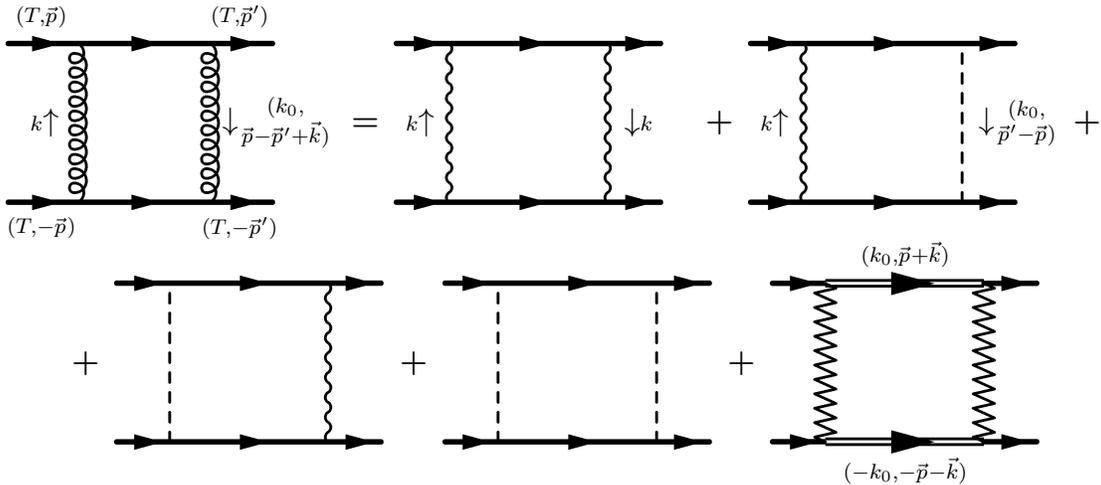
The ultraviolet behaviour of this graph is mimicked in NRQFT by a four-fermion
exchange given by the vertex $\ii c_2=\frac{-\ii g^4}{24 \pi^2
  M^2}=\calO(t^0,y^0)$ of the Lagrangean (\ref{nrlagr}), which using the
rescaling rules is seen to be $\calO(v)$.

The Feynman rules (\ref{pupvertex}) give that the $\Au\Au$-diagram is of order
$\e^v$ with a leading loop integral contribution (similar to \cite[fl.\ 
(32)]{BenekeSmirnov})
\begin{equation}
  \label{vertex2uu}
  \int\dedk \frac{1}{k_0^2-\kv^2}\;
  \frac{1}{k_0^2-\kv^2}\;\frac{1}{T+k_0-\frac{\pv^2}{2M}}
  \;\frac{1}{T-k_0-\frac{\pv^2}{2M}}\;\;.
\end{equation}
The diagram is expected to be zero to all orders since the ultrasoft gluons do
not change the quark momenta and therefore the scattering takes place only in
the forward direction, $\pv=\pv^\prime$. Upon employing the on-shell condition
for potential quarks, $T=\frac{\pv^2}{2M}$ to leading order, it indeed vanishes
as no scale is present. Since $T-\frac {\pv^2}{2M}\sim Mv^4\ll |\kv|\sim Mv$
(and $k_0\sim Mv^2$) in the potential r\'egime, this is a legitimate expansion.
The $\Au\Ap$ and $\Ap\Au$ contributions ($\calO(\frac{1}{v}\e^v)$) are zero for
the same reason. The lowest order contribution to the $\Ap\Ap$ graph
($\calO(\frac{1}{v^2})$) is
\begin{equation}
  \label{vertex2pp}
  \int\dedk \frac{1}{\kv^2-\ii\epsilon}\;\frac{1}{(\pv-\pv^\prime+\kv)^2
  -\ii\epsilon}\; \frac{1}{T+k_0-\frac{(\kv+\pv)^2}{2M}+\ii\epsilon}\;
  \frac{1}{T-k_0-\frac{(\kv+\pv)^2}{2M}+\ii\epsilon}\;\;.
\end{equation}
In the light of the discussion at the end of Sect.\ \ref{philosophy}, it is
most consistent to perform the $k_0$ integration by dimensional regularisation,
using $\int\deintdim{d}{k}=\int\deintdim{\sigma}{k_0}
\deintdim{d-\sigma}{\kv}$, $\sigma\to1$ \cite[Chap.\ 4.1]{Collins}. Split
dimensional regularisation was introduced by Leibbrandt and Williams
\cite{LeibbrandtWilliams} to cure the problems arising from pinch singularities
in non-covariant gauges. Appendix \ref{app:splitdimreg} shows that in the case
at hand, it has the same effect as closing the $k_0$-contour and picking the
quark propagator poles prior to using dimensional regularisation in $d-1$
Euclidean dimensions. To achieve $\calO(v^1)$ accuracy, one must also consider
one insertion (\ref{insertions}) at the potential gluon lines, giving rise to a
contribution
\begin{eqnarray}
  \label{vertex2pp2}
  && \int\dedk \frac{1}{\kv^2-\ii\epsilon}\;
  \frac{1}{(\pv-\pv^\prime+\kv)^2-\ii\epsilon}\;
  k_0^2\;\Big(\;\frac{1}{\kv^2-\ii\epsilon}\;+\;
  \frac{1}{(\pv-\pv^\prime+\kv)^2
  -\ii\epsilon}\;\Big)\;\times\nonumber \\
  &&\;\;\;\;\;\;\;\;\;\;\;\;\;
  \times\;\frac{1}{T+k_0-\frac{(\kv+\pv)^2}{2M}+\ii\epsilon}\;
  \frac{1}{T-k_0-\frac{(\kv+\pv)^2}{2M}+\ii\epsilon}\;\;.
\end{eqnarray}
The $k_0$ integration is na{\ia}vely linearly divergent, and hence closing the
contour is not straightforward. As App.\ \ref{app:splitdimreg} demonstrates,
split dimensional regularisation circumvents this problem.  The sum of both
contributions (\ref{vertex2pp}/\ref{vertex2pp2}),
\begin{equation}
  \label{vertex2ppresult}
  \frac{\ii}{8\pi t}\;\frac{M+T}{\sqrt{y}}\;\Big(\frac{2}{4-d}-
  \gamma_\mathrm{E}-\ln\frac{-t}{4\pi\mu^2}\Big)\;\;,
\end{equation}
agrees with \cite[fl.\ (31)]{BenekeSmirnov} when one keeps in mind that in that
reference, heavy particle external lines were normalised relativistically,
while a non-relativistic normalisation was chosen here. Also, this article uses
the $MS$ rather than the $\overline{MS}$ scheme. Near threshold, the scale is
set by the total threshold energy $4\pi\mu^2=4(M+T)^2$.

The soft gluon part is to lowest order ($\calO(v^{-1})$ because of one soft
blob) given by
\begin{equation}
  \label{vertex2ss}
  \int\dedk\frac{1}{k_0^2-\kv^2+\ii\epsilon}\;\frac{1}{k_0^2-(\pv-\pv^\prime+
\kv)^2+\ii\epsilon}\;\frac{1}{k_0+\ii\epsilon}\;\frac{1}{-k_0+\ii\epsilon}\;\;,
\end{equation}
which corresponds to \cite[fl.\ (33)]{BenekeSmirnov}. Now, split dimensional
regularisation must be used if no ad-hoc prescription for the pinch singularity
at $k_0=0$ is to be invoked. That the pinch is accounted for by potential gluon
exchange and hence must be discarded, agrees with the intuitive argument that
zero four-momentum scattering in QED is mediated by a potential only, and no
retardation or radiation effects occur. On the other hand, the model Lagrangean
contains three marginal couplings as seen at the end of Sect.\ \ref{rescaling},
which may give finite contributions as energies and momenta of the scattered
particles go to zero. Although the prescription and the result from split
dimensional regularisation coincide in the present case as demonstrated at the
end of Sect.\ \ref{philosophy} and in App.\ \ref{app:splitdimreg}, this may not
hold in general. The result to $\calO(v^1)$ exhibits another collinear
divergence,
\begin{equation}
  \label{vertex2ssresult}
  \frac{-\ii}{4\pi^2 t}\;\Big(\frac{2}{4-d}-\gamma_\mathrm{E}-
  \ln\frac{-t}{4\pi\mu^2}\Big)\;+\;\frac{\ii}{24\pi^2M^2}\;\Big[1+ \frac{2y}{t}
   \;\Big(\frac{2}{4-d}-\gamma_\mathrm{E}-\ln\frac{-t}{4\pi\mu^2}\Big)\Big]
   \;\;,
\end{equation}
and agrees with the one given by Beneke and Smirnov \cite[fl.\ 
(36)]{BenekeSmirnov}. The second term comes from insertions and multipole
expansions to achieve $\calO(v)$ accuracy.

It is easy to see that the power counting proposed works.  As expected, the
potential diagram is $\sqrt{y}\propto v$ stronger that the leading soft
contribution, and $t \sqrt{y}\propto v^3$ stronger than the four-fermion
interaction.

In conclusion, the proposed NRQFT Lagrangean reproduces the result for the
planar graph of the relativistic theory \emph{only if} the soft gluon and the
soft quark are accounted for: The four-fermion contact interaction produces
just a $\frac{1}{M^2}$-term, graphs containing ultrasoft gluons were absent,
and the potential gluon (\ref{vertex2ppresult}) gave no $\calO(y^0)$
contribution.  This shows the necessity of soft quarks and gluons. The coupling
strength of the $\Phis\As\Phip$ vertex is also seen to be identical to the
other vertex coupling strengths, $g$.

The planar fourth order correction to two quark production (Fig.\ \ref{graph3})
was also compared to the result of \cite{BenekeSmirnov}, and is correctly
accounted for when the Feynman rules proposed above are used to $\calO(v^1)$.
\begin{figure}[!htb]
  \vspace*{2ex}
  \begin{center}
    \feynbox{100\unitlength}{
    \begin{fmfgraph*}(90,60)
      \fmfstraight \fmfleft{i1} \fmfright{o2,o1}
      \fmf{ghost,width=thick,tension=1.5,label=$\fs (2T,,\vec{0})$}{i1,v1}
      \fmfdot{v1} \fmf{fermion,width=thick,tension=0.5}{v1,v2,v3}
      \fmf{fermion,width=thick,tension=0.5}{v1,v4,v5}
      \fmf{fermion,width=thick,tension=2}{v5,o2}
      \fmf{fermion,width=thick,tension=2}{v3,o1} \fmflabel{$\fs (T,\pv)$}{o1}
      \fmflabel{$\fs (T,-\pv)$}{o2} \fmffreeze \fmf{gluon,width=thin,label=$\fs
        {\dis\uparrow} k$,label.side=right}{v4,v2}
      \fmf{gluon,width=thin,label=$\fs {\dis\downarrow}
        l$,label.side=left}{v3,v5}
    \end{fmfgraph*}
    }\hq\bf{=}\hq \feynbox{90\unitlength}{ \begin{fmfgraph*}(90,60)
      \fmfstraight \fmfleft{i1} \fmfright{o2,o1}
      \fmf{ghost,width=thick,tension=2}{i1,v1} \fmfdot{v1}
      \fmf{fermion,width=thick,tension=0.5}{v1,v2,v3}
      \fmf{fermion,width=thick,tension=0.5}{v1,v4,v5}
      \fmf{fermion,width=thick,tension=2}{v5,o2}
      \fmf{fermion,width=thick,tension=2}{v3,o1} \fmffreeze
      \fmf{photon,width=thin}{v4,v2} \fmf{dashes,width=thin}{v3,v5}
    \end{fmfgraph*} }
  \hq\bf{+}\hq \feynbox{90\unitlength}{
    \begin{fmfgraph*}(90,60) \fmfstraight \fmfleft{i1} \fmfright{o2,o1}
      \fmf{ghost,width=thick,tension=2}{i1,v1} \fmfdot{v1}
      \fmf{fermion,width=thick,tension=0.5}{v1,v2,v3}
      \fmf{fermion,width=thick,tension=0.5}{v1,v4,v5}
      \fmf{fermion,width=thick,tension=2}{v5,o2}
      \fmf{fermion,width=thick,tension=2}{v3,o1} \fmffreeze
      \fmf{dashes,width=thin}{v4,v2} \fmf{dashes,width=thin}{v3,v5}
    \end{fmfgraph*} }\hq\bf{+}\hq\feynbox{90\unitlength}{
    \begin{fmfgraph*}(90,60) \fmfstraight \fmfleft{i1} \fmfright{o2,o1}
      \fmf{ghost,width=thick,tension=2}{i1,v1} \fmfdot{v1}
      \fmf{fermion,width=thick,tension=0.5}{v1,v2}
      \fmf{fermion,width=thick,tension=0.5}{v1,v4}
      \fmf{heavy,width=thin,tension=0.5}{v2,v3}
      \fmf{heavy,width=thin,tension=0.5}{v4,v5}
      \fmf{fermion,width=thick,tension=2}{v5,o2}
      \fmf{fermion,width=thick,tension=2}{v3,o1} \fmffreeze
      \fmf{zigzag,width=thin}{v4,v2} \fmf{zigzag,width=thin}{v3,v5}
    \end{fmfgraph*}}
\end{center}
\caption{\figlabel{graph3}\sl The non-vanishing contributions to the planar
  fourth order correction to two quark production. Diagrams with insertions or
  four-point interactions not displayed.}
\end{figure}
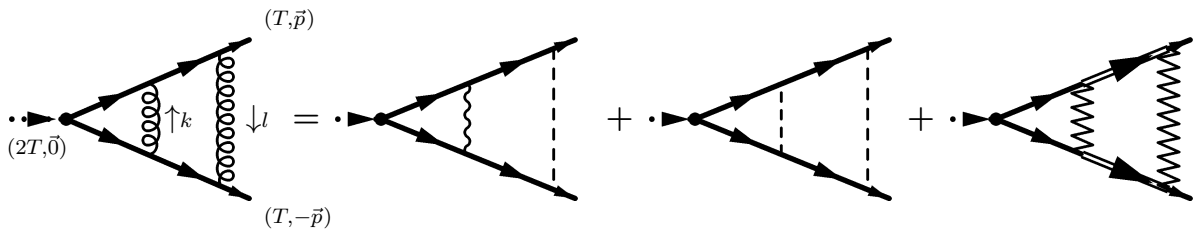

\section{Conclusions and Outlook}
\seclabel{conclusions}
\setcounter{equation}{0}

The objective of this article was a simple presentation of the ideas behind
explicit velocity power counting in dimensionally regularised NRQFT. It started
with the identification of three different r\'egimes of scale for on-shell
particles in NRQFT from the poles in the non-relativistic propagators. This
leads in a natural way to the existence of a new quark field and a new gluon
field in the soft scaling r\'egime $E\sim |\pv|\sim Mv$. In it, quarks are
static and gluons on shell, and HQET becomes a sub-set of NRQCD. Neither of the
five fields in the three r\'egimes should be thought of as ``physical
particles''. Rather, they represent the ``true'' quark and gluon in the
respective r\'egimes as the infrared-relevant degrees of freedom. None of the
r\'egimes overlap.  An NRQFT Lagrangean has been proposed which leads to the
correct behaviour of scattering and production amplitudes. It establishes
explicit velocity power counting which is preserved to all orders in
perturbation theory. The reason for the existence of such a Lagrangean, once
dimensional regularisation is chosen to complete the theory, was elaborated
upon in a simple example: the non-commutativity of the expansion in small
parameters with dimensionally regularised integrals.

Due to the similarity between the calculation of the examples in the work
presented here and in \cite{BenekeSmirnov}, one may get the impression that the
Lagrangean presented is only a simple re-formulation of the threshold
expansion. Partially, this is true, and a future publication \cite{hgpub4} will
indeed show the equivalence of the two approaches to all orders in the
threshold and coupling expansion. A list of other topics to be addressed there
contains: the straightforward generalisation to NRQCD; a proof whether the
particle content outlined above is not only consistent but complete,
i.e.\ that no new fields (e.g.\ an ultrasoft quark) or ``exceptional''
r\'egimes arise; an investigation of the influence of soft quarks and gluons on
bound state calculations in NRQED and NRQCD; a full list of the various
couplings between the different r\'egimes and an exploitation of their
relevance for physical processes. The formal reason why double counting between
different r\'egimes and especially between soft and ultrasoft gluons does not
occur, a derivation of the way soft quarks couple to external sources, and the
r\'ole of soft gluons in Compton scattering deserve further attention, too.

I would like to stress that the diagrammatic threshold expansion derived here
allows for a more automatic and intuitive approach and makes it easier to
determine the order in $\sqrt{-y}\propto v$ to which a certain graph
contributes. On the other hand, the NRQFT Lagrangean can easily be applied to
bound state problems. As the threshold expansion of Beneke and Smirnov starts
in a relativistic setting, it may formally be harder to treat bound states
there. Indeed, I believe that even if one may not be able to prove the
conjectures of the one starting from the other, both approaches will profit
from each other in the wedlock of NRQFT and threshold expansion.

\section*{Acknowledgments} 
It is my pleasure to thank J.-W.\ Chen, D.\ B.\ Kaplan and M.\ J.\ Savage
for stimulating discussions. The work was supported in part by a Department of
Energy grant DE-FG03-97ER41014.


\begin{appendix}
\section{Some Details on Split Dimensional Regularisation}
\seclabel{app:splitdimreg}
\setcounter{equation}{0}

This appendix presents the part of the calculations in the examples of Sect.\ 
\ref{bsexamples} which makes use of split dimensional regularisation as
introduced by Leibbrandt and Williams \cite{LeibbrandtWilliams}. In its
results, split dimensional regularisation agrees with other methods to compute
loop integrals in non-covariant gauges, such as the non-principal value
prescription \cite{LeeNyeo}, but two features make it especially attractive: It
treats the temporal and spatial components of the loop integrations on an equal
footing, and no recipes are necessary. Rather, it uses the fact that, like in
ordinary integration, the axioms of dimensional regularisation \cite[Chap.\ 
4.1]{Collins} allow to split the integration into two separate integrals:
\begin{equation}
  \label{eq:splitdimreg}
  \int\deintdim{d}{k}=\int\deintdim{\sigma}{k_0}\deintdim{d-\sigma}{\kv}
\end{equation}
Both integrations can be performed consecutively, and the limit $\sigma\to 1$
can -- if finite -- be taken immediately, because the integration over the
spatial components of the loop momentum in (\ref{eq:splitdimreg}) is still
regularised in $d-1$ dimensions. Finally, the limit $d\to 4$ is taken at the
end of the calculation.

Equation (\ref{vertex2pp}) contains the simplest $k_0$ sub-integral:
\begin{equation}
  \label{eq:vertex2ppsdr}
  \int\deintdim{\sigma}{k_0}
  \frac{1}{k_0+T-\frac{(\kv+\pv)^2}{2M}+\ii\epsilon}\;
  \frac{1}{-k_0+T-\frac{(\kv+\pv)^2}{2M}+\ii\epsilon}
\end{equation}
Using standard formulae for dimensional regularisation in Euclidean space
\cite[App.\ B]{Ramond}, the result is finite as $\sigma\to 1$:
\begin{eqnarray}
  \label{eq:vertex2ppsdrres}
  && \int\deintdim{\sigma}{k_0}
    \frac{1}{[T-\frac{(\kv+\pv)^2}{2M}+\ii\epsilon]^2 -k_0^2}
  = -\frac{\Gamma[1-\frac{\sigma}{2}]}{(4\pi)^{\sigma/2} \Gamma[1]} \;
    \bigg(-\Big[T-\frac{(\kv+\pv)^2}{2M}+\ii\epsilon\Big]^2
    \bigg)^{\frac{\sigma}{2}-1}\to\nonumber\\
  &&\to -\frac{\ii}{2}\;\Big(T-\frac{(\kv+\pv)^2}{2M}+\ii\epsilon\Big)^{-1}
    \;\;\mbox{ as }\sigma\to 1
\end{eqnarray}
It is no surprise that closing the contour produces the same result, because
for any finite integral, the answer of all regularisation methods have to
coincide. The integral over the spatial components of the loop momentum is now
straightforward.

The potential gluon diagram with one insertion at a gluon leg
(\ref{vertex2pp2}) yields a split dimensional integral which diverges linearly
in $k_0$, so that na{\ia}ve contour integration is not legitimate.
\begin{equation}
  \label{eq:vertex2pp2sdr}
  \int\deintdim{\sigma}{k_0}
    \frac{k_0^2}{[T-\frac{(\kv+\pv)^2}{2M}+\ii\epsilon]^2 -k_0^2}
  \to - \frac{\ii}{2}\;\Big(T-\frac{(\kv+\pv)^2}{2M}\Big)\;\;\mbox{ as }
     \sigma\to 1  
\end{equation}
To arrive at this result, the numerator was re-written
as $(k_0^2- (T-\frac{(\kv+\pv)^2}{2M})^2) \;+\;(T-\frac{(\kv+\pv)^2}{2M})^2$.
Its first term cancels the denominator, yielding an integral without scale
which therefore vanishes in dimensional regularisation. The second term has
been calculated in (\ref{eq:vertex2ppsdrres}). The integral over the spatial
components of the loop momentum provides again no complications, leading to
(\ref{vertex2ppresult}).

Finally, it was already shown at the end of Sect.\ \ref{philosophy} that
dimensional regularisation discards pinch singularities encountered in contour
integrations. This is validated again by looking at the split dimensional
integral for $k_0$ in the soft gluon contribution (\ref{vertex2ss}),
\begin{equation}
  \label{eq:vertex2sssdr}
  \int\deintdim{\sigma}{k_0} \frac{1}{k_0^2-a^2}\;
  \frac{1}{k_0^2-b^2}\;
  \frac{1}{-k_0^2}\;\;,
\end{equation}
where $a^2:=\kv^2-\ii\epsilon$ and $b^2:=(\pv-\pv^\prime+\kv)^2-\ii\epsilon$. 
After combining denominators, the resulting integral is simple:
\begin{eqnarray}
  -2\;\frac{\Gamma[3-\frac{\sigma}{2}]}{(4\pi)^\frac{\sigma}{2} \Gamma[3]}\;
    \int\limits^1_0 \deint{}{x}\deint{}{y} x\;\Big(-a^2 (1-x)-b^2 xy
    \Big)^{\frac{\sigma}{2}-3}\to
    \frac{\ii}{2}\; \frac{1}{a^2-b^2}\;
    \Big(\frac{1}{a^3}-\frac{1}{b^3}\Big)\;\;\mbox{ as } \sigma\to 1 \;\;.
\end{eqnarray}
This agrees with the result of Beneke and Smirnov \cite[fl.\ 
(34)]{BenekeSmirnov} who use contour integration and drop the contribution from
the pinch singularity. The integral over the spatial components of the loop
momentum provides again no unfamiliar complications, leading to
(\ref{vertex2ssresult}).

\end{appendix}  


\end{fmffile}
\end{document}